\documentclass[aps, prl, twocolumn]{revtex4-2}
\usepackage{amsmath}
\usepackage[colorlinks, linkcolor=red, anchorcolor=green, citecolor=blue]{hyperref}

\usepackage{graphicx}
\usepackage{epsfig}
\usepackage{bm}
\usepackage{color}
\usepackage{amsmath}
\usepackage{dcolumn}
\usepackage{enumitem}
\usepackage{orcidlink}

\usepackage{amssymb}
\begin{document}

\title{Attractor for 1+1D viscous hydrodynamics with general rapidity distribution}
\begin{abstract}
Attractor, supposed to be one of the possible answer for the early applicability of hydrodynamics in the evolution with different initial conditions, has attracted great attention to the fast decreasing of degrees of freedom in heavy ion collision. We found attractor behaviors for 1+1D viscous hydrodynamics with general rapidity distribution based on MIS theory. Meanwhile, we also observe that a rapid expansion in the fluid velocity is essential for a rapid early time attractor.
\end{abstract}
\author{Shile Chen\orcidlink{0000-0002-3874-5564}}\email{csl2023@tsinghua.edu.cn}
\affiliation{Department of Physics, Tsinghua University, Beijing 100084, China}
\author{Shuzhe Shi\orcidlink{0000-0002-3042-3093}}\email{shuzhe-shi@tsinghua.edu.cn}
\affiliation{Department of Physics, Tsinghua University, Beijing 100084, China}
\maketitle

\emph{Introduction.} --- 
Relativistic heavy-ion collisions, pursued at the Relativistic Heavy-Ion Collider (RHIC) and at the Large Hadron Collider (LHC), allow systematic studies of a color-deconfined phase of matter -- the Quark-Gluon Plasma(QGP). One of the breakthroughs in theoretical relativistic heavy-ion physics has been the realization of the great success of numerical hydrodynamic simulations in understanding and predicting measurements of the collective behavior of the observed hadrons (see e.g. Refs.~\cite{Gale:2013da, Schenke:2010rr, Shen:2014vra, Schenke:2010rr}). 
Such numerical simulations show the applicability of hydrodynamics after an unexpected short time after the initial collision. The hydrodynamics theory is generally believed to apply for systems near local equilibrium, whereas the initial condition of the relativistic heavy-ion collisions is far from equilibrium. The reason of fast hydrodynamization in relativistic heavy-ion collisions remains a puzzle, and the hydro attractor~\cite{Heller:2015dha} is plausibly the answer.

The M\"uller--Isreal--Stewart (MIS) theory of hydrodynamics is commonly adopted in the numerical simulations. It contains conservation of energy-moment, equation of state, and relaxation of the shear stress tensor to its Navier--Stokes limit,
\begin{align}
    &\mathcal{D}_\mu T^{\mu\nu} = 0\,,\label{eq:conservation}\\
    &\tau_\pi \Delta_{\alpha\beta}^{\mu\nu} u^\lambda \mathcal{D}_\lambda \pi^{\alpha\beta} = -\pi^{\mu\nu} + \eta\,\sigma^{\mu\nu}\,,\label{eq:shear}
\end{align}
where $\mathcal{D}_\mu$ is the covariant derivative, $\sigma^{\mu\nu} = 2 \Delta_{\alpha\beta}^{\mu\nu} \mathcal{D}^\alpha u^{\beta}$ the shear tensor, $\eta = C_\eta \frac{\varepsilon+P}{T}$ the shear viscosity, and $\tau_\pi = \frac{C_\tau}{T}$ the relaxation time. 
The spatial projection operators are defined as $\Delta^{\mu\nu} = g^{\mu\nu} - u^\mu u^\nu$, $\Delta^{\mu\nu}_{\alpha\beta} = \frac{1}{2}\Delta^{\mu}_{\alpha} \Delta^{\nu}_{\beta} + \frac{1}{2}\Delta^{\mu}_{\beta} \Delta^{\nu}_{\alpha} - \frac{1}{3}\Delta^{\mu\nu} \Delta_{\alpha\beta}$.
The shear stress tensor is orthogonal to the fluid velocity,
$u_\mu \pi^{\mu\nu} = 0$, and traceless $\Delta_{\mu\nu}\pi^{\mu\nu} = 0$.
By taking the conformal limit, the energy density is given by $\varepsilon = \frac{3}{\pi^2}T^4$, the pressure is $P = \varepsilon/3$, and the bulk pressure vanishes, so that the decomposition of the stress tensor is $T^{\mu\nu} = (\varepsilon+P) u^\mu u^\nu - P\,g^{\mu\nu} + \pi^{\mu\nu}$.
The $\mathcal{N}$ Super Yang--Mills theory~\cite{Bhattacharyya:2007vjd} gives that $C_\eta = \frac{1}{4\pi}$ and $C_\tau = \frac{2-\ln2}{2\pi}$. 

Heller and Spalinski~\cite{Heller:2015dha} found the first attractor solution for viscous hydrodynamics in $(0+1)$ dimensions, which applies for systems that are boost-invariant in the longitudinal direction and homogeneous in transverse direction. 
In $(0+1)$ D, there are only two independent hydrodynamic variables, the energy density and one of the diagonal shear viscous tensor, and they depend only on the proper time, but not the longitudinal or transverse coordinates.
It was found, in~\cite{Heller:2015dha}, that by introducing the dimensionless variables $w \equiv \tau \,T$ and
$f \equiv \tau \frac{\partial_\tau w}{w} = 1 + \tau\frac{\partial_\tau T}{T}$, Eqs.~\eqref{eq:conservation} and~\eqref{eq:shear} can be combined into a single-variable equation that contains up to first-order derivatives,
\footnote{Note that some factors are different from Eq. (9) of ~\cite{Heller:2015dha} since \eqref{eq:shear} only keeps up to first  order terms.}
\begin{align}
    w\,f\frac{\mathrm{d}f}{\mathrm{d}w} + 4f^2 + \Big(\frac{w}{C_\tau} - \frac{20}{3}\Big)f = 
    \frac{4\,C_\eta}{9\,C_\tau} - \frac{8}{3} + \frac{2\,w}{3\,C_\tau}\,.\label{eq:attractor}
\end{align}
Ref.~\cite{Heller:2015dha} found that solutions of \eqref{eq:attractor} with different initial conditions converge to a universal curve, called \emph{attractor}, within $w \sim C_\tau$.
Noting that $w/C_\tau = \tau/\tau_\pi$ is the ratio between proper time and the temperature-dependent relaxation time, the attractor behavior implies rapid convergence of the hydrodynamic modes.

Other attractor solutions are found in viscous hydrodynamics that obeys Gubser symmetry~\cite{Behtash:2017wqg, Denicol:2018pak, Dash:2020zqx}, anisotropic hydrodynamics with Bjorken-like~\cite{Strickland:2017kux}
 expansions and even for Boltzmann equation with relaxation time kernel~\cite{Strickland:2018ayk, Almaalol:2020rnu}. 
Attractors in higher-order viscous hydro~\cite{Jaiswal:2019cju}, 
Boltzmann equation for ultrarelativistic gas with hard-sphere interaction ~\cite{Denicol:2019lio} are also found. Such attractor solutions correspond to systems with space-time dependence highly constrained by symmetry properties. 
Attractors for hydrodynamics with general spatial dependence is not found yet. 
Attractors for hydro solution that violates homogeneity in the transverse plane is also explored in Ref.~\cite{Romatschke:2017acs, Ambrus:2021sjg}, but large-time convergence that is independent of the spacial coordinates was not observed. 
Meanwhile, it was speculated~\cite{Kurkela:2019set} that the longitudinal expansion, rather than interaction, is essential for attractor.
In the current paper, we study the attractors for $(1+1)$ D viscous hydrodynamics that remain homogeneous in the transverse plane but breaks the boost-invariance. 

One may generalize the Bjorken flow to a hydro solution with non-trivial rapidity dependence by performing a ``temporal shift''~\cite{Shi:2022iyb, Chen:2023vrk}. 
For the generalized solution, one may find that the attractor solution obtained in~\cite{Heller:2015dha} remains applicable after redefining the quantities in a Lorentz-invariant manner, i.e., one should replace the proper-time derivative by the co-moving derivative, $\mathcal{D}_\tau \to u^\mu \mathcal{D}_\mu$, and the proper time shall be replaced by the inverse of the expansion rate, $\tau \to 1/\theta \equiv 1/(\mathcal{D}_\mu u^\mu)$. 
That is, the dimensionless quantities in the attractor shall be replaced as,
\begin{align}
    \tilde{w} \equiv T/\theta,\qquad
    \varphi \equiv 1+\frac{u^\mu \partial_\mu T}{\theta\,T}\,.
    \label{eq:novel_attractor}
\end{align}
Such a generalization can be justified as follows. From the energy-momentum conservation equation~\eqref{eq:conservation}, one may derive that $1+ 3 \frac{u^\mu \mathcal{\partial}_\mu T}{T\,\theta} = \frac{\pi^{\mu\nu} \sigma_{\mu\nu}}{2\,s\,T\,\theta}$, which further leads to  $\varphi_\mathrm{NS}(\tilde{w}) = \frac{2}{3}-\frac{4C_\eta}{9\tilde{w}}$ when taking the Navier--Stokes theory that $\pi^{\mu\nu} = \eta\,\sigma^{\mu\nu}$. Therefore, quantities defined in Eq.~\ref{eq:novel_attractor} can at least converge to $\varphi_\mathrm{NS}(\tilde{w})$.
Non-trivially, by numerically solving the MIS viscous hydro equations with general initial rapidity distribution, we observe that \eqref{eq:novel_attractor} exhibits (early) attractor solution that follows \eqref{eq:attractor} by replaces that $f\to\varphi$ and $w\to\tilde{w}$. We note that such quantities are similar to those defined in~\cite{Romatschke:2017acs}.

\begin{figure}
    \centering
    \includegraphics[width=0.45\textwidth]{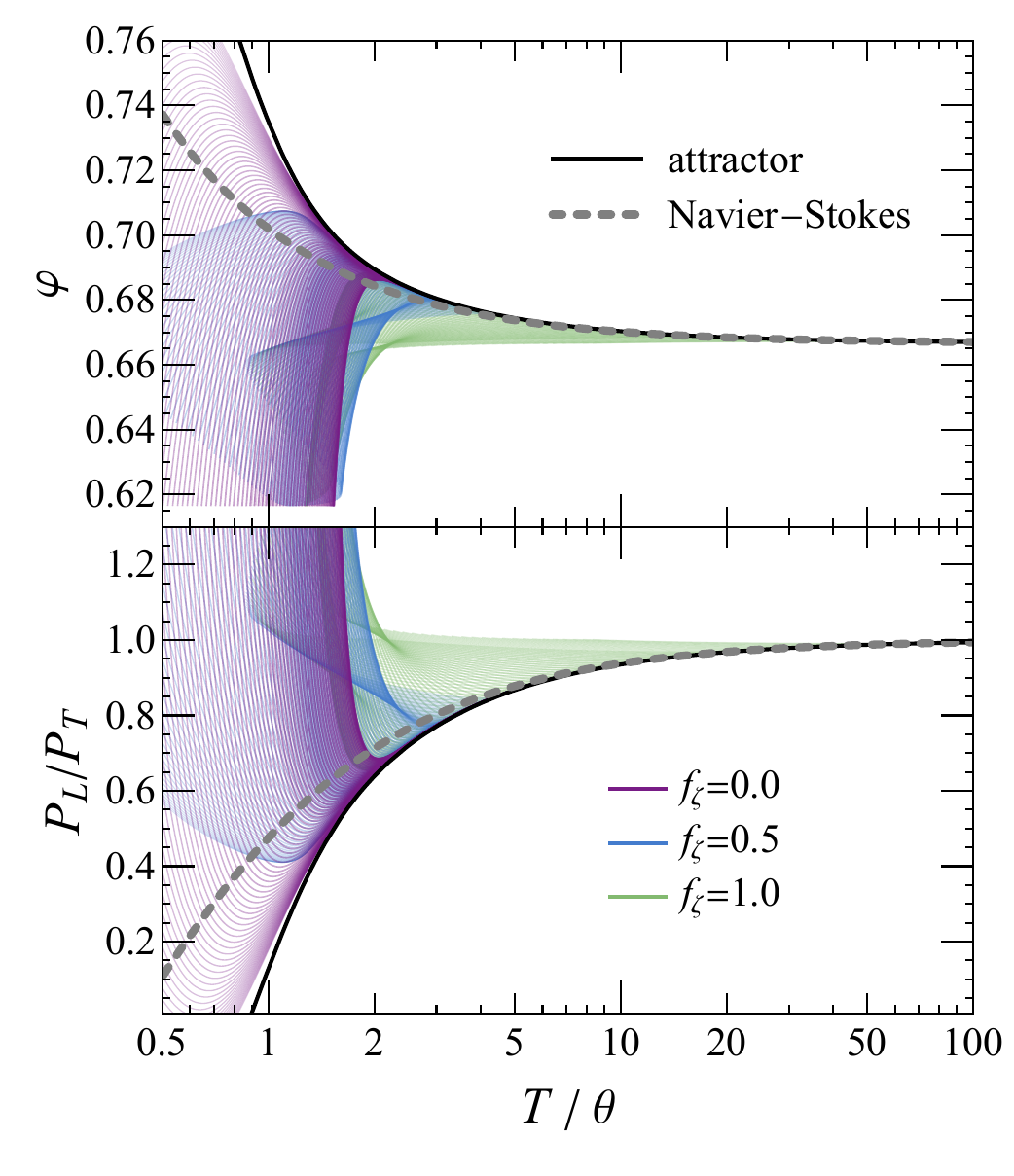}
    \caption{Novel attractor quantity (upper) and pressure anisotropy (lower) versus the scaled co-moving time.
    In all four panels, thin curves in purple, blue, and green correspond to three different hydro profile with initial flow parameter $f_\zeta=0$, $f_\zeta=0.5$, and $f_\zeta=1$, respectively, whereas different curves in the same color are from different rapidity slides in the same hydro profile. 
    For comparison, black thick solid curves correspond to attractor solutions whereas gray dotted curves are the 0+1D solution of Navier--Stokes hydro.
    \label{fig:zeta}}
\end{figure}

\begin{figure*}
    \centering
    \includegraphics[width=0.45\textwidth]{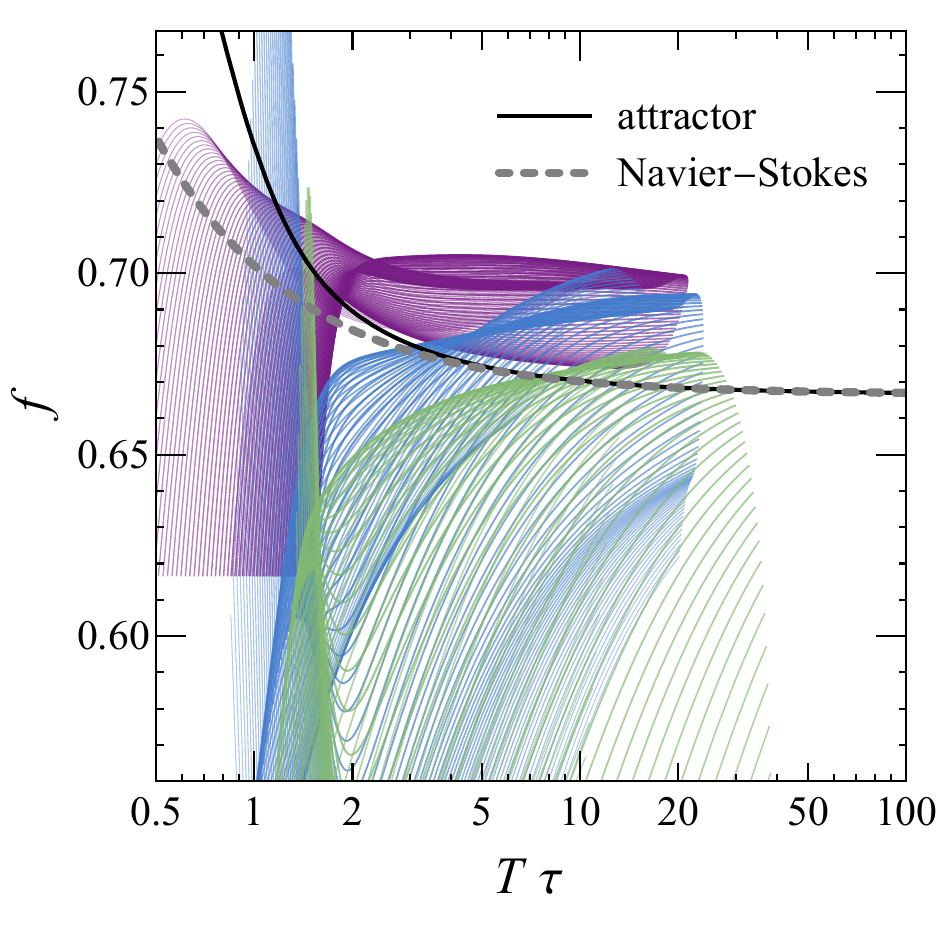}
    \includegraphics[width=0.45\textwidth]{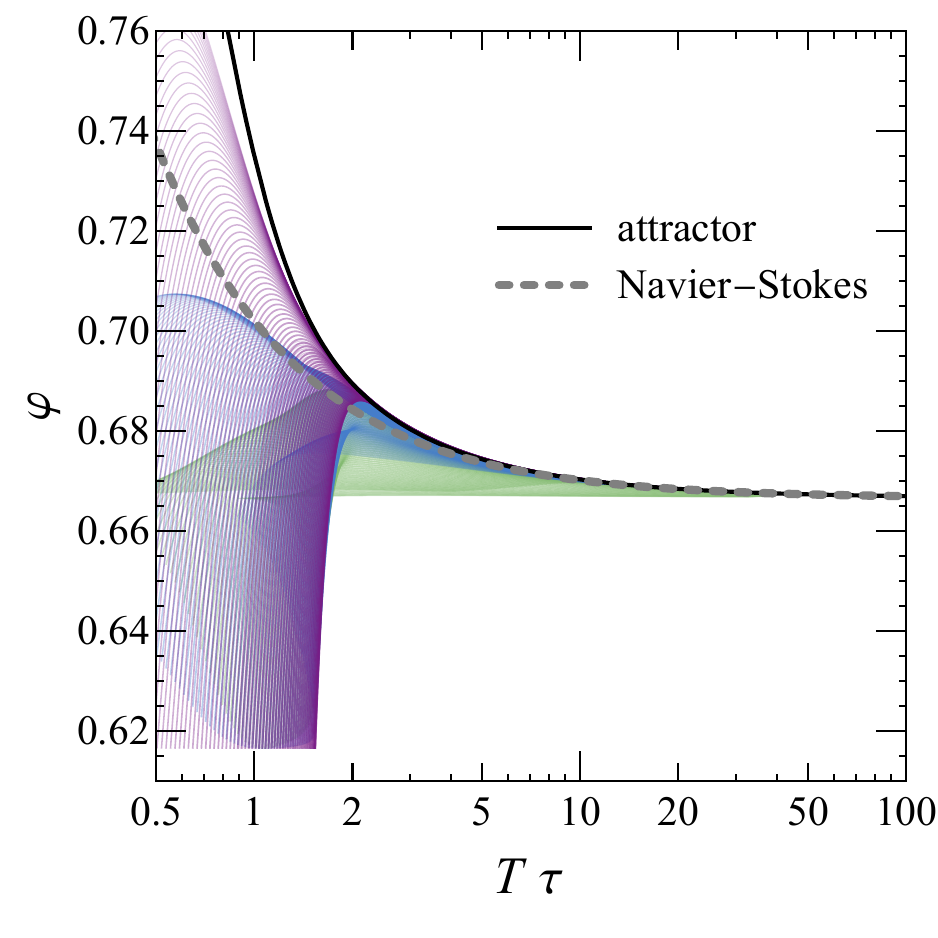}
    \caption{
    Bjoken-based (left) and novel (right) attractor quantities versus scaled proper time.
    Parameter settings and color coding are the same as Fig.~\protect{\ref{fig:zeta}}.
    \label{fig:tau}}
\end{figure*}

\begin{figure*}
    \centering
    \includegraphics[width=0.45\textwidth]{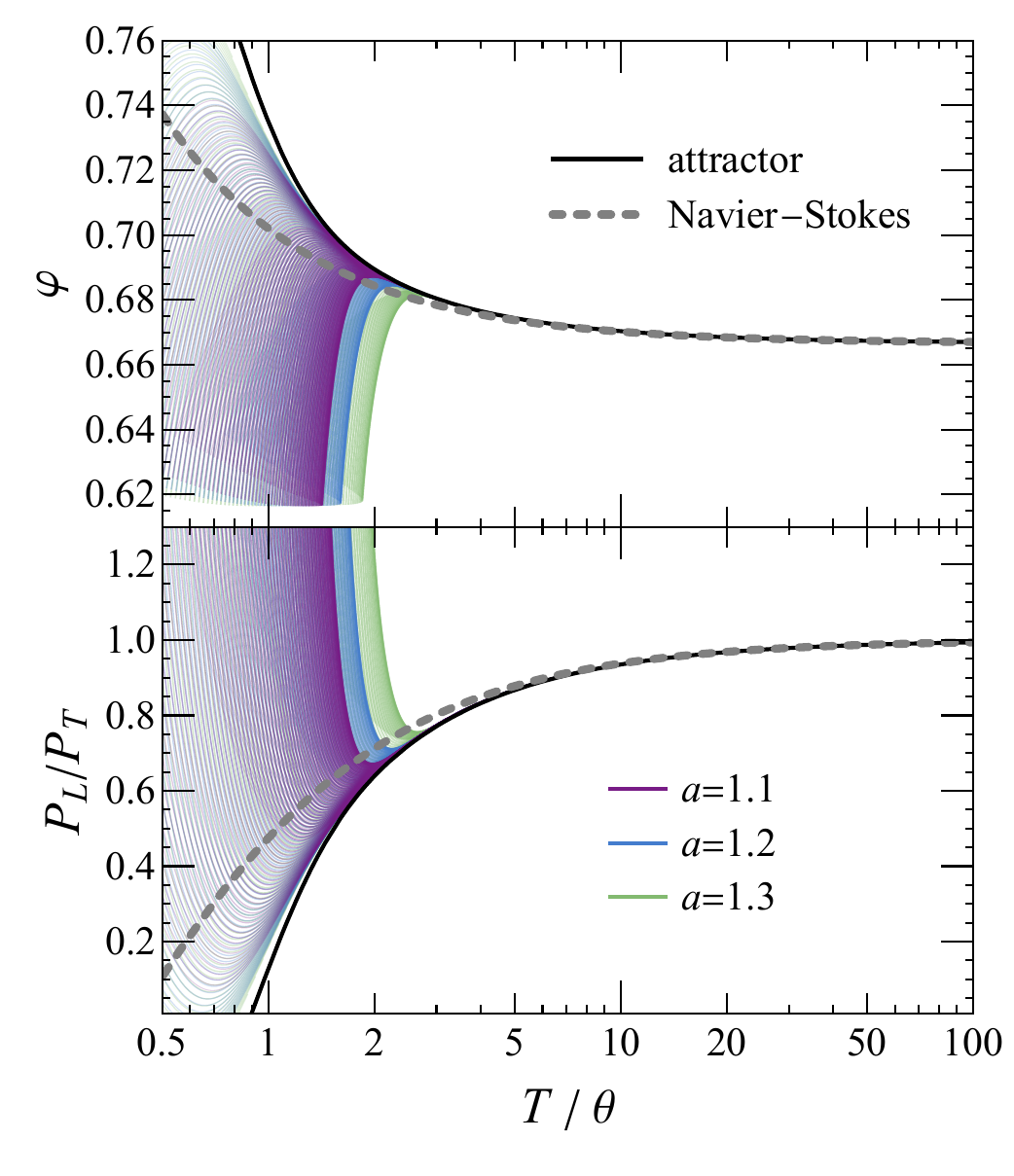}
    \includegraphics[width=0.45\textwidth]{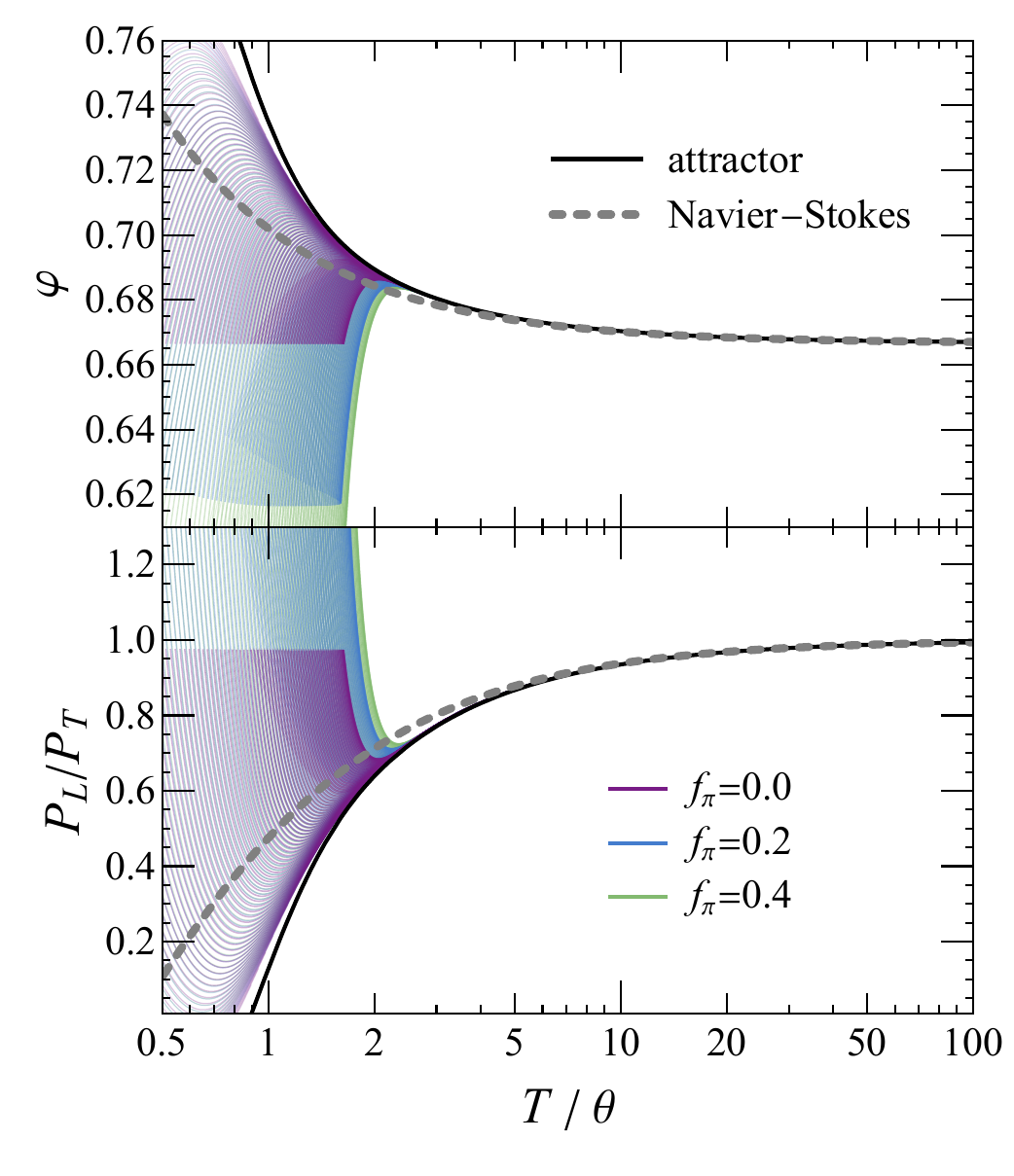}
    \caption{Same as Fig.~\protect{\ref{fig:zeta}} but for the comparison of initial parameters $a\in\{1.1,1.2,1.3\}$ (left) and $f_\pi\in\{0.0,0.2,0.4\}$ (right).
    \label{fig:a-pi}}
\end{figure*}

\vspace{5mm}
\emph{Numerical solution.} --- 
In a system that is homogeneous in the transverse plane, one has $u^x = u^y = 0$ and $\partial_x = \partial_y = 0$. 
We further let $u^\tau \equiv \cosh\frac{\zeta}{2}$ and $u^\eta \equiv \frac{1}{\tau}\sinh\frac{\zeta}{2}$, so that the normalization condition is satisfied.
The hydro equations~\eqref{eq:conservation} and~\eqref{eq:shear} become
\begin{align}
\begin{split}
0 =\,& \partial_\tau
    \Big(\frac{2\varepsilon}{3}\cosh\zeta +\frac{\varepsilon}{3}+\pi^{\tau\tau}\Big)
+
    \frac{4\varepsilon}{3\tau}\cosh\zeta
\\&+
    \partial_\eta\Big(\frac{2\varepsilon}{3\tau}\sinh\zeta 
    + \pi^{\tau\eta}\Big)
    + \tau\pi^{\eta\eta}
    +\frac{\pi^{\tau\tau}}{\tau},
\end{split}\label{eq:conservation_2}\\
\begin{split}
0 =\,& \Big(\frac{3}{\tau} + \partial_\tau \Big)\Big(\frac{2\varepsilon}{3\tau}\sinh\zeta +\pi^{\tau\eta}\Big) \\&+ 
    \partial_\eta \Big(\frac{2\varepsilon}{3\tau^2}\cosh\zeta -\frac{\varepsilon}{3\tau^2}+\pi^{\eta\eta}\Big)\,,
\end{split}
\end{align}
\begin{align}
\begin{split}
0=\,&u^\lambda \partial_\lambda \pi^{\tau\tau} 
    - \pi^{\tau\tau} \coth\frac{\zeta}{2} u^\lambda \partial_\lambda\zeta
    - \frac{\eta\,\sigma^{\tau\tau}-\pi^{\tau\tau}}{\tau_\pi}\,.
\end{split}\label{eq:shear_2}
\end{align}
Different components of the $\pi$ tensor are connected with each other when implementing the orthogonal relation.

In this work, we initialize the hydro profile by a parametrization inspired by an asymmetric analytical solution~\cite{Shi:2022iyb}
\begin{align}
\frac{\varepsilon}{\varepsilon_i} =\,& 
    (t_0 + a\,\tau_i e^\eta)^{\frac{2}{3}(a^{-2}-2)} (t_0 + \tau_i e^{-\eta}/a)^{\frac{2}{3}(a^{2}-2)}\,,\\
\zeta=\,& 
    f_\zeta \ln\frac{t_0 e^{-\eta} + a\,\tau_i }{t_0 e^\eta + \tau_i/a}\,,\\
\pi^{\eta\eta} =\,& f_\pi \frac{\varepsilon}{\tau_i^2}\,.
\end{align}
$\tau_i$ is the initial proper time, $t_0$ is of the time unit and controls the rapidity plateau, $1/\sqrt{2} < a < \sqrt{2}$ controls the level of asymmetry, and $f_\zeta$ and $f_\pi$ are additional parameters alters the initial profile from the analytical solution. Taking $f_\zeta = 1$ and $f_\pi=0$, as well as turning of shear viscosity $C_\eta=0$, the solution would return to the analytical solution found in~\cite{Shi:2022iyb}. We fix $t_0=0.1 \tau_i$ and $\varepsilon_i = \tau_i^{-4}$ and vary $a$, $f_\zeta$, and $f_\pi$ and check attractor behavior.
We note that such initialization parametrization is for the convenience of controlling different components, and the conclusion does not rely on it.

We begin by exploring the influence of initial flow velocity. We take $a=1.2$, $f_\pi=0.2$, and $f_\zeta=\{0,0.5,1\}$, and numerically solve Eqs.~(\ref{eq:conservation_2} -- \ref{eq:shear_2}) with the corresponding initial profile. After obtaining the space-time dependent temperature and velocity, we construct $\tilde{w}$ and $\varphi$ according to \eqref{eq:novel_attractor}, as well as the pressure anisotropy $\frac{P_L}{P_T}=\frac{\varepsilon+3\pi^{x}_{x}}{\varepsilon-3\pi^{x}_{x}/2}$.
We plot them for different rapidity slides and present the results in Fig.~\ref{fig:zeta}. For all rapidity slides and for all values of $f_\zeta$, all curves eventually approach the attractor solution, i.e., solution of \eqref{eq:attractor} but with the replacement that $f\to\varphi$ and $w\to\tilde{w}$. Meanwhile, the attractor of pressure anisotropy is given by $A^{P}(\tilde{w}) = \frac{3-4\varphi(\tilde{w})}{2\varphi(\tilde{w})-1}$, see e.g., \cite{Strickland:2017kux}.
For assessment of the attractor speed, we also show results from the Navier--Stokes theory, which takes $\pi^{\mu\nu} = \eta\sigma^{\mu\nu}$ rather than solving the relaxation time equation~\eqref{eq:shear} and gives that $\varphi_\mathrm{NS}(\tilde{w}) = \frac{2}{3}-\frac{4C_\eta}{9\tilde{w}}$ and $A^{P}_\mathrm{NS}(\tilde{w}) = \frac{3-4\varphi_\mathrm{NS}(\tilde{w})}{2\varphi_\mathrm{NS}(\tilde{w})-1}$. It is apparent that the hydro solution converges to the attractor solution faster than the convergency between attractor and Navier--Stokes solution when $f_\zeta=0$ or $0.5$, whereas late time attractor is observed when $f_\zeta=1$. From~\cite{Shi:2022iyb} one can learn that analytical solution, corresponding to $f_\zeta=1$, expands slower than the Bjorken flow, i.e., $f_\zeta=0$. We conclude that the expansion with the speed of a Bjorken flow --- or at least close enough --- is essential for the early time attractor. 

For comparison, we also present the Bjorken-based and the novel attractor quantities versus the temperature scaled proper time, see Fig.~\ref{fig:tau}. It is clear that no attractor behavior is observed for the Bjorken-based quantity. While there seems to be some attractor behavior in $\varphi(w)$, we emphasize that $\varphi(\tilde{w})$ is a better one. In addition to the theoretical advantage that $\tilde{w}$ is invariant under Lorentz boost along any direction, the phenomenological advantage is more apparent in the $\tilde{w} \in [1,2]$ sector of the blue curves, and the narrower white region between the purple curves and the attractor at early time.

After exploring the effect of initial flow, we move on to study the asymmetry and initial shear. We fix $f_\zeta=0.2$, and take $f_\pi=0.2$ and $a\in\{1.1,1.2,1.3\}$ in the former, and $a=1.2$ and $f_\pi\in\{0,0.2,0.4\}$ in the latter. Results are shown in Fig.~\ref{fig:a-pi}. In general, we observe that hydro solution approaches to the attractor more rapidly in more symmetric system (with $a$ closer to unity) or with less initial shear stress tensor (smaller $f_\pi$). Early time attractor behavior is observed in all these parameter sets.


\vspace{5mm}
\emph{Summary and Outlooks.} In this work, we numerically solve the 1+1 dimensional viscous hydrodynamic equation with general initialization. We propose a novel attractor that  replaces the quantities of the attractor found in Ref.~\cite{Heller:2015dha} into their Lorentz-invariant version, and we find that the hydro solution approaches to the novel attractor solution. We also observe that the initial expansion rate that is close to Bjorken flow is essential for the emergency of an early-time attractor.

\vspace{5mm}
\emph{Acknowledgement.} We thank Dr. Lipei Du for very helpful discussion. This work is supported by Tsinghua University under grant Nos. 53330500923 and 100005024.

\end{document}